\documentclass[aps,prl,twocolumn,showpacs,superscriptaddress]{revtex4-1}
\pdfoutput=1

\usepackage{graphicx}
\usepackage{dcolumn}
\usepackage{color}

\usepackage{eucal}
\usepackage{amssymb}

\usepackage{amsmath,amsthm}

\begin{document}
\title{Giant adsorption of microswimmers: duality of shape asymmetry and wall curvature.}

\author{Adam Wysocki, Jens Elgeti, and Gerhard Gompper}
\affiliation{Theoretical Soft Matter and Biophysics, 
Institute of Complex Systems and Institute for Advanced Simulation,
Forschungszentrum J\"ulich, 52425 J\"ulich, Germany}

\date{\today}
\begin{abstract}
The effect of shape asymmetry of microswimmers on their adsorption
capacity at confining channel walls is studied by a simple dumbbell model. 
For a shape polarity of a forward-swimming cone, like the 
stroke-averaged shape of a sperm, extremely long wall retention times are found, 
caused by a non-vanishing component of the propulsion force pointing steadily 
into the wall, which grows exponentially with the self-propulsion 
velocity and the shape asymmetry. 
A direct duality relation between shape asymmetry and wall curvature 
is proposed and verified. Our results are relevant for the design 
microswimmer with controlled wall-adhesion properties. 
In addition, we confirm that pressure in active systems is strongly sensitive 
to the details of the particle-wall interactions.
\end{abstract}



\maketitle

Boundaries dominate biological process on all scales. 
On the microscopic scale, motile organism may bump into various 
obstactles and boundaries, such as liquid-gas 
or liquid-liquid interfaces, elastic cell membranes or solid walls. 
A universal behavior is the accumulation of microswimmers at boundaries 
\cite{lauga2009rpp,elgeti2015rpp}. 
Aside from physico-chemical effects \cite{tuson2013sm}, such as 
van der Waals forces, two main mechanisms have been suggested to explain 
the wall accumulation, hydrodynamic interactions (HI) \cite{berke2008prl,drescher2011pnas} 
and excluded-volume (or steric) forces \cite{elgeti2009epl,li2009prl}. 
The importance of HI on accumulation is still a subject of debate. 
However, recent experiment in quasi-2D microfluidic channels indicate that 
surface scattering of sperm and {\em Chlamydomonas} at lateral boundaries 
is dominated by steric forces with multiple flagellar contacts 
\cite{kantsler2013pnas,denissenko2012pnas},  
in strong contrast to sperm confined in a 3D channel, where HI seems to
dominate adhesion \cite{elgeti10bj}.

\begin{figure}[t]
\centering
\includegraphics[width=0.7\columnwidth]{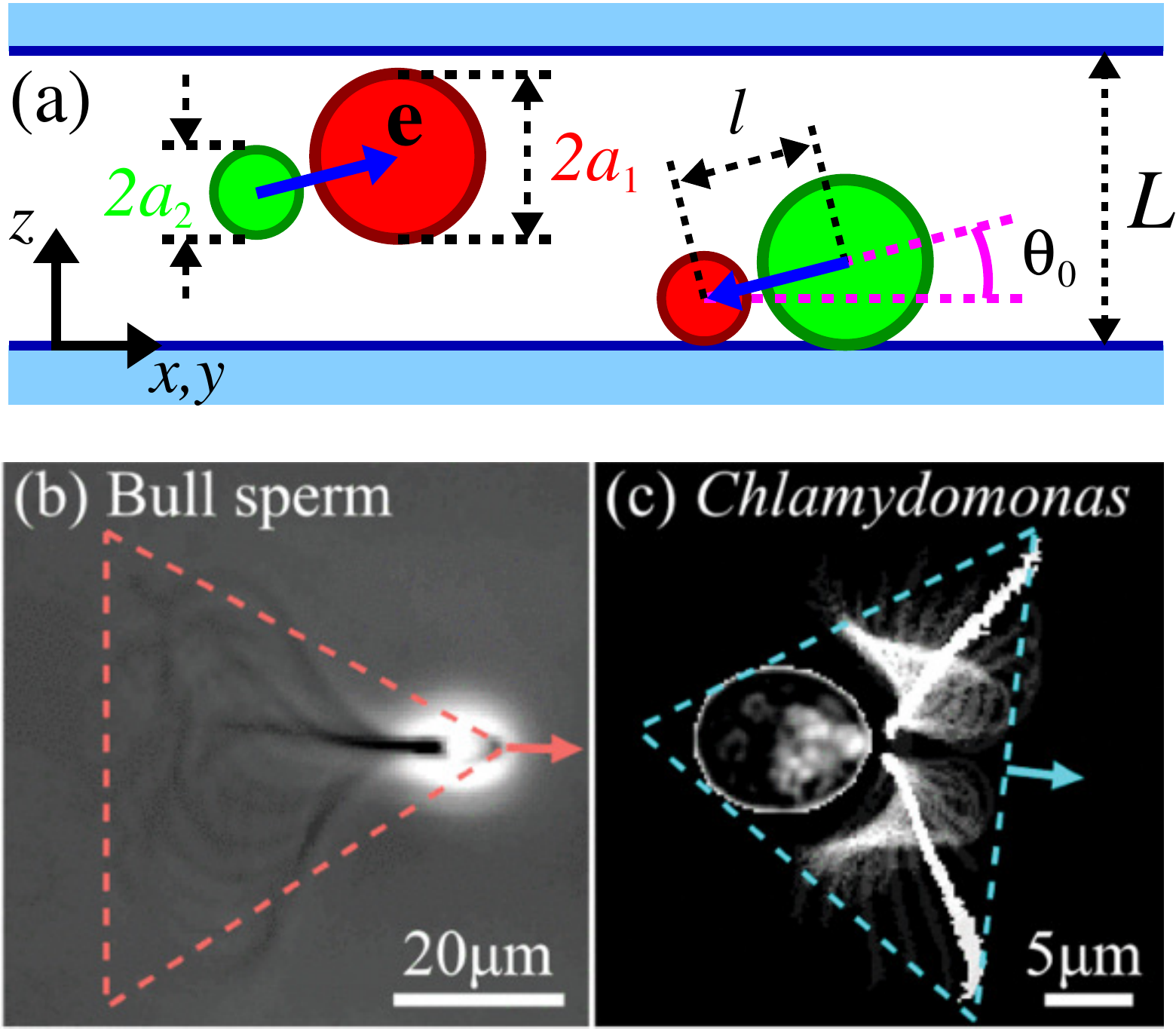}
\caption{\label{f:sketch}(color online) (a) Sketch of the dumbbell model
of asymmetric microswimmers in confinement. The swimmer propel 
along its instantaneous orientation $\mathbf{e}$ with velocity $V$ 
in a channel of height $L$. Left: a antipolar particle ($\theta_0>0$); 
right: an polar particle ($\theta_0<0$). (b,c) Experimental images. 
(b) Superimposed phase-contrast micrographs of swimming bull sperm; 
the cell mimics a forward-swimming cone (polar shape). 
(b) A {\it Chlamydomonas} alga, confined to quasi-2D motion, 
resembles a backward-swimming triangle (antipolar shape). 
Reprinted with permission from Ref.~\cite{wensink2014pre}.}
\end{figure}

Most theoretical studies of simple model swimmers, both in bulk and in 
confinement, have considered so far cells with a symmetric body shape, in particular, 
rods \cite{elgeti2009epl,li2009prl} or spherical particles 
\cite{elgeti2013epl,lee2013njp}. 
In reality, however, cells usually do not exhibit such high symmetry, 
and the stroke-averaged shape of sperm or {\em Chlamydomonas} rather
resembles a forward or a backward swimming cone, respectively 
\cite{kantsler2013pnas,wensink2014pre,denissenko2012pnas}, see also Fig.~\ref{f:sketch}(b,c). 
This raises the question how a broken for-aft symmetry of the particle 
shape alters the wall accumulation of cells.

In order to elucidate shape effects on effective adsorption, we neglect HI and 
study a generic model of an active Brownian dumbbell  
with unequal bead sizes, see Fig.~\ref{f:sketch}(a). 
Our simulation results show that swimmers 
with a sperm-like (polar) shape exhibit huge wall trapping times 
due to a nonvanishing component of the propulsion force directed steadily 
toward the wall, thus resulting in a restricted rotational movement. 
The trapping times increase exponentially with the shape asymmetry 
$\theta_0$ and the propulsion strength $V$ and could, for realistic 
parameters of $\theta_0$ and $V$, exceed trapping times due to near-field hydrodynamic 
forces \cite{drescher2011pnas,spagnolie2014sm,schaar2014arXiv}. 
In contrast, microswimmers with {\em Chlamydomonas}-like (antipolar) 
shape behave similarly to symmetric rod-like particles. 

Both in a natural environment and in microfluidic devices 
\cite{kaehr2009lc,denissenko2012pnas,wioland2013prl}, microswimmers usually do not swim in straight, 
but rather in curved or branching microchannels. Therefore, the influence 
of surface curvature on accumulation of microswimmers 
\cite{teeffelen2009sm,fily2014sm,takagi2014sm,spagnolie2014sm,vladescu2014prl,guidobaldi2015bmf} 
is of great interest. Based on the analysis of an asymmetric particle near 
a flat boundary, we predict a direct duality relation between the effect 
of shape asymmetry and surface curvature on accumulation. 
For example, a polar microswimmer close to a flat wall behaves similarly to 
an apolar particle near a concave surface (e.g. a cavity). 
This is of high relevance for the design microswimmers with controlled
wall-adhesion properties. 

We model the microswimmer as a self-propelled Brownian dumbbell. 
The dumbbell consists of two spheres with radii $a_1$ and $a_2$ connected 
by a rigid rod of length $l$, see Fig.~\ref{f:sketch}(a). 
Its orientation is characterized by a unit vector $\mathbf{e}$ directed 
along the axis from sphere $2$ to sphere $1$. 
The equation of motion for the swimmer's center (midpoint between the sphere centers) is then
\begin{equation}
\dot{\mathbf{r}}=V\mathbf{e}+\mathbf{\Xi}^{-1}\mathbf{F}_w+\boldsymbol{\eta},
\label{eq:BDtrans}
\end{equation}
where $V$ is the bare propulsion velocity, $\mathbf{\Xi}$ is the translational friction tensor, $\mathbf{F}_w$ is the steric force due to swimmer-wall interaction and $\boldsymbol{\eta}$ 
is a random velocity. The particle is confined in a channel of height $L$ along the $z$-direction. 
The sphere $\alpha\in\{1,2\}$ interacts with the walls via a screened Coulomb potential $U_{\alpha}$ 
\footnote{The sphere $\alpha\in\{1,2\}$ interacts with the lower wall via
\begin{equation}
\frac{U_{\alpha}}{k_BT}=10\frac{\exp{\left[-\kappa(z_{\alpha}-a_{\alpha})\right]}}{\kappa(z_{\alpha}-a_{\alpha})},
\label{eq:yukawa}
\end{equation}
and equivalently with the upper wall, where $z_{\alpha}=\mathbf{r}_{\alpha}\hat{\mathbf{z}}$ 
is the $z$-coordinate of sphere $\alpha$. Strong screening is achieved by using $\kappa a=10$.
}
with a large inverse screening length $\kappa$ and thus resembles a hard-sphere. 
The total dumbbell-wall force is $\mathbf{F}_w=\mathbf{F}_1+\mathbf{F}_2$ with 
$\mathbf{F}_{\alpha}=-\nabla_{\mathbf{r}_{\alpha}}U_{\alpha}$ where $\mathbf{r}_{\alpha}$ is the position of 
sphere $\alpha$. 
The Gaussian white-noise velocity $\boldsymbol{\eta}$ obeys 
$\langle\boldsymbol{\eta}(t)\boldsymbol{\eta}(t')\rangle=2k_BT\mathbf{\Xi}^{-1}\delta(t-t')$, 
where $k_BT$ is the thermal energy scale and $\mathbf{\Xi}=\gamma_{\parallel}\mathbf{e}\mathbf{e}+\gamma_{\perp}(\mathbf{I}-\mathbf{e}\mathbf{e})$ 
is the translational friction tensor with the friction coefficients $\gamma_{\parallel}$ 
and $\gamma_{\perp}$ for motions parallel and perpendicular to $\mathbf{e}$, respectively. 
The orientation evolves according to
\begin{equation}
\dot{\mathbf{e}}=(\boldsymbol{T}_w/\gamma_r+\boldsymbol{\xi})\times\mathbf{e},
\label{eq:BDrot}
\end{equation}
where $\boldsymbol{\xi}$ is a Gaussian white-noise vector with 
$\langle\boldsymbol{\xi}(t)\boldsymbol{\xi}(t')\rangle=2D_r\mathbf{I}\delta(t-t')$ 
and $D_r=k_BT/\gamma_r$ is the rotational diffusion coefficient. 
The torque due to the wall interaction is $\boldsymbol{T}_w=\boldsymbol{T}_1+\boldsymbol{T}_2$ 
with 
$\boldsymbol{T}_1=(\mathbf{r}_1-\mathbf{r})\times\mathbf{F}_1=l(\mathbf{e}\times\mathbf{F}_1)/2$ 
and $\boldsymbol{T}_2=-l(\mathbf{e}\times\mathbf{F}_2)/2$. 
We solve Eqs.~(\ref{eq:BDtrans}) and (\ref{eq:BDrot}) numerically using standard methods 
\cite{loewen1994pre}. 

Dimensionless numbers characterising the system are the Peclet number $Pe=V/(lD_r)$, 
which is the ratio of the swimming persistence length $V/D_r$ to the rod length $l$, 
and the shape asymmetry parameter $\sin{(\theta_0)}=(a_1-a_2)/l$, which is 
$\sin{(\theta_0)}<0$ for polar (sperm-like) and $\sin{(\theta_0)}>0$ 
for antipolar ({\it Chlamydomonas}-like) microswimmers. 
Wherever appropriate, we choose realistic parameters similar to that of 
{\it Escherichia coli}  
\footnote{Using realistic parameters for {\it Escherichia coli} \cite{drescher2011pnas},
we set length scales to $l=8$ $\mu$m and $a=a_1+a_2=1$ $\mu$m, 
diffusion constants to $D_{\parallel}=0.149$ $\mu$m$^2/$s, $D_{\perp}=0.135$ $\mu$m$^2/$s, 
and $D_r=0.032$ s$^{-1}$. We vary $Pe$ by changing $V$ up to $60$ $\mu$ms$^{-1}$.
}. 
We increase swimming velocity up to $Pe=234$; for comparison, {\it Escherichia coli} 
achieve $Pe\approx120$ \cite{drescher2011pnas}, while {\it Chlamydomonas} and bull sperm 
reach only $Pe\approx25-50$ due to the large active rotational diffusion 
\cite{drescher2011pnas,tung2015prl}. 
We consider only small asymmetry, $|\sin{(\theta_0)}|\approx|\theta_0|\le0.125$. 
Note that $|\sin{(\theta_0)}|\approx0.5$ for sperm and {\it Chlamydomonas} 
\cite{wensink2014pre}, see also Fig.~\ref{f:sketch}(b,c). 
Furthermore, we apply a weak confinement with channel height $L=10l$.
\begin{figure}[t]
\centering
\includegraphics[width=1\columnwidth]{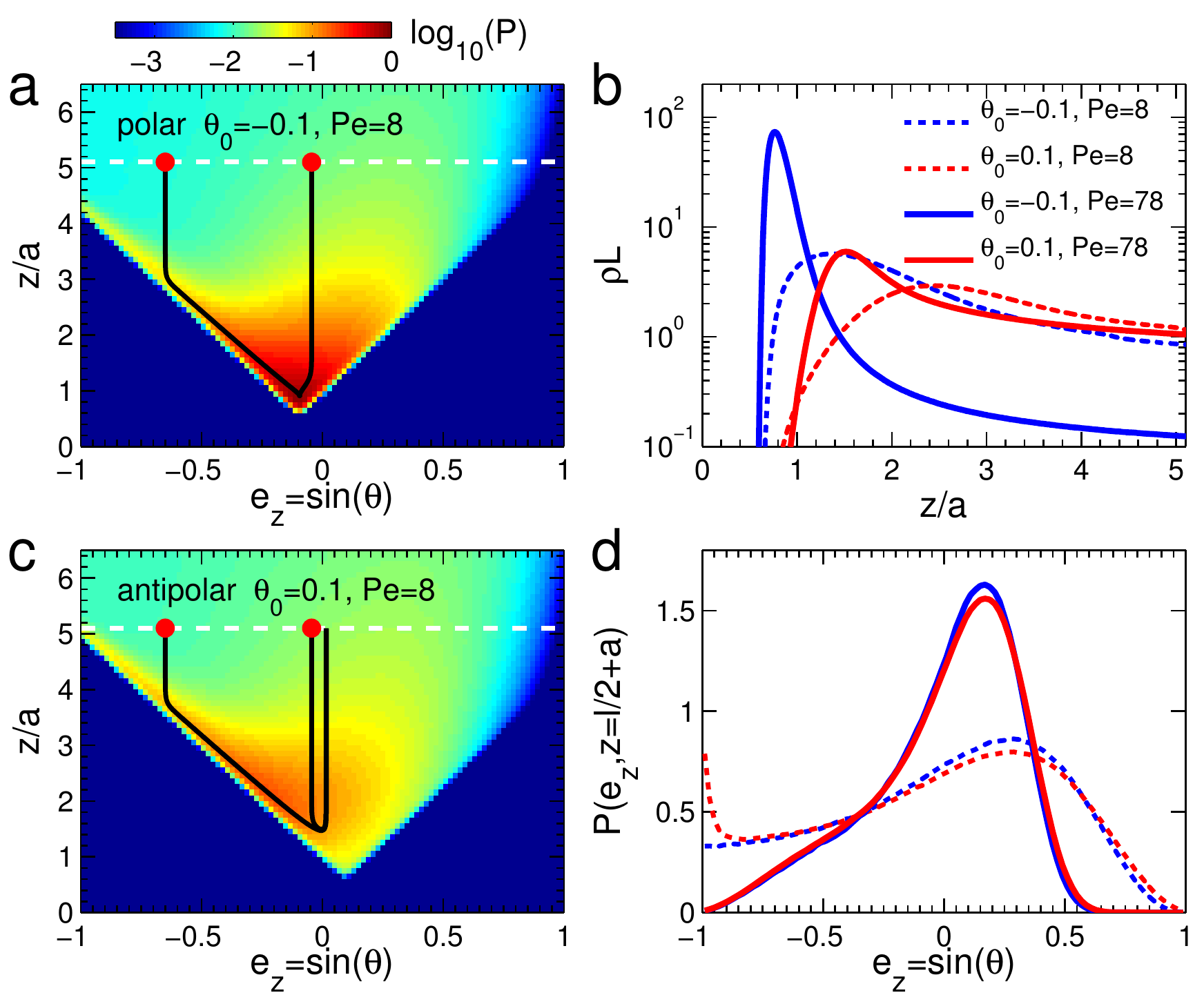}
\caption{\label{f:rho_ez_rz}(color online) (a,c) Logarithmically color-coded plots 
of the probability density function (PDF) $P(e_z,z)$ at moderate activity $Pe=8$. 
(a) $P(e_z,z)$ for a polar ($\theta_0<0$) and (c) an antipolar swimmer ($\theta_0>0$). 
Deterministic trajectories are indicated by solid lines.
A polar swimmer is highly localised at $(e_z,z)=(\sin{(\theta_0)},a/2)$. 
(b) The corresponding orientation-averaged PDF $\rho(z)=\int_{-1}^{1}P(e_z,z)\,\mathrm{d}e_z$. 
(d) PDF of the orientation $P(e_z,z\approx\delta)$ at the threshold of steric interactions 
($\delta=l/2+a$), see dashed line in (a,c). 
Swimmers leave the wall region increasingly parallel with increasing $Pe$.} 
\end{figure}

\begin{figure}[t]
\centering
\includegraphics[width=1\columnwidth]{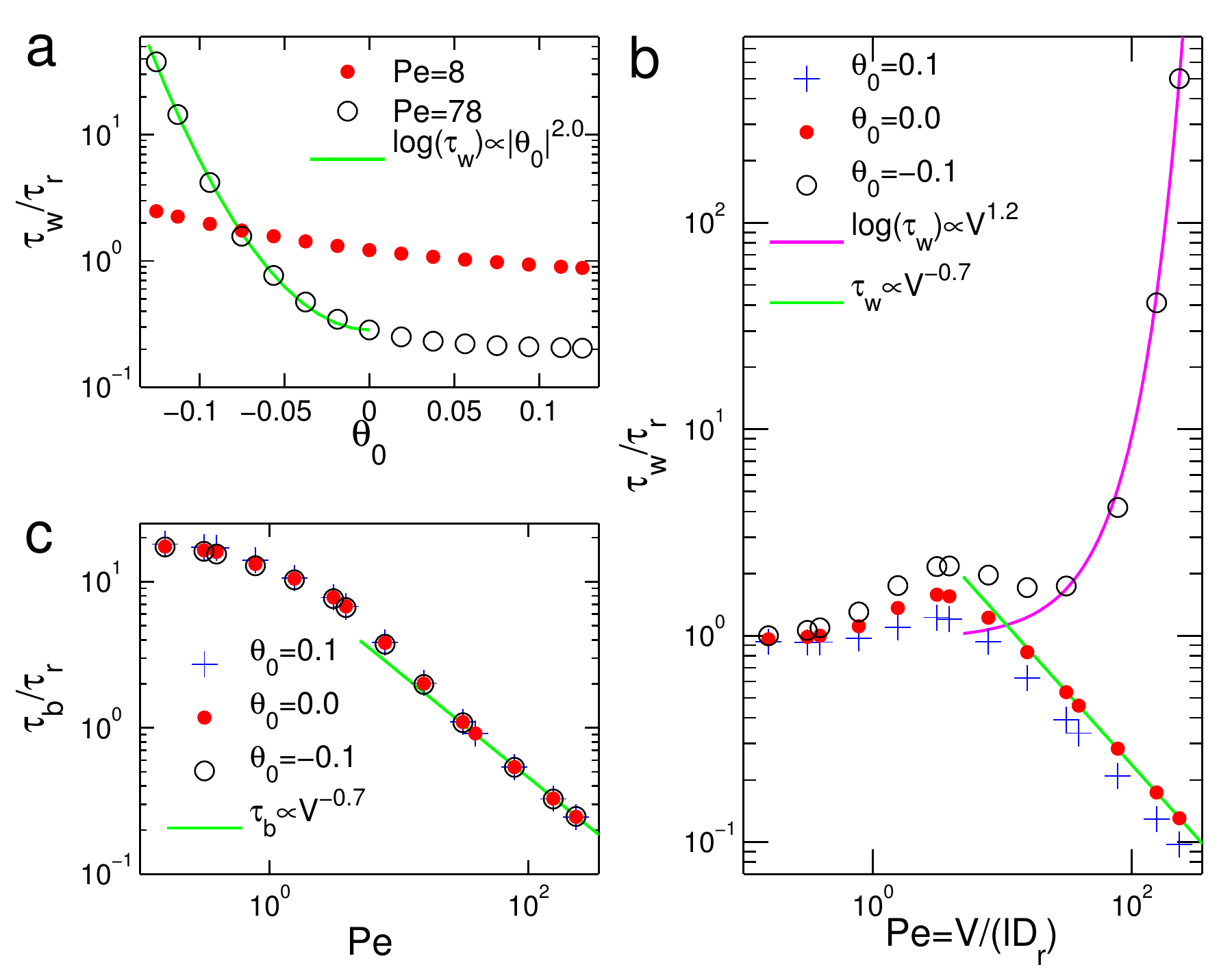}
\caption{\label{f:t_wall_t_bulk}(color online) (a) Mean time $\tau_w$ a swimmer remains 
within the wall region ($z\in[0,\delta]$ or $z\in[L-\delta,L]$) as function of the 
shape asymmetry $\theta_0$; time is normalized by $\tau_r=1/(2D_r)$. 
(b) $\tau_w$ versus $Pe$. For polar swimmers $\tau_w$ grows exponentially with $V$, 
while for apolar and antipolar swimmers $\tau_w\propto V^{-0.7}$. 
(c) Mean time $\tau_b$ a swimmer remains within the bulk region ($z\in[\delta,L-\delta]$) 
versus $Pe$. $\tau_b$ is independent of $\theta_0$ and $\tau_b\propto V^{-0.7}$ for $Pe\gg L/l=10$.}
\end{figure}

There are universal features in the behavior of an elongated microswimmer confined 
inside a channel. The swimmer performs a persistent random walk 
within the bulk region; when it encounters a wall, a torque, caused by steric interactions, 
leads to approximately parallel alignment with the wall; finally, the swimmer can 
escape from the 
boundary when its orientation, as a result of rotational diffusion, points slightly 
away from the boundary \cite{elgeti2009epl,li2009prl}. 
Generally, swimmers are increasingly localized near the wall with increasing activity, 
as can be seen from the density profile $\rho(z)$ in Fig.~\ref{f:rho_ez_rz}(b). 
Shape asymmetry changes the behavior dramatically. Polar swimmers are much more strongly 
adsorbed as their antipolar counterpart, see Fig.~\ref{f:rho_ez_rz}(b). Moreover, 
polar particles point persistently toward the wall at an angle prescribed by the 
body shape, see high probability density at $(e_z,z)=(\sin{(\theta_0)},a/2)$ in 
Fig.~\ref{f:rho_ez_rz}(a) and compare to the probability density function $P(e_z,z)$ 
of a antipolar swimmer in Fig.~\ref{f:rho_ez_rz}(c). A high probability density near 
the boundary is tantamount to a large wall retention time. As can be seen 
in Fig.~\ref{f:t_wall_t_bulk} and will be discussed in more detail below, 
the retention time of the particles indeed increases very rapidly 
with increasing activity and shape polarity. 

In order to understand the giant wall accumulation of polar microswimmers, and in particular 
the huge trapping times, we examine Eqs.~(\ref{eq:BDtrans}) and (\ref{eq:BDrot}) 
with the focus on the two relevant variables, the coordinate along the surface normal $z$ 
and the orientation angle $\theta$, see Fig.~\ref{f:sketch}, which implies  
\begin{align}
\label{eq:BDsimple1}
\dot{z}     &=V\sin{(\theta)}+F_w/\gamma  +\eta,\\
\label{eq:BDsimple2}
\dot{\theta}&=-D_r\tan{(\theta)}+T_w/\gamma_r+\xi.
\end{align}
The first term on the RHS of Eq.~(\ref{eq:BDsimple2}) is peculiar for 
rotational diffusion in 3D \cite{raible2004aomc,schaar2014arXiv} 
and can be neglected for $Pe\gg1$.
The noise obeys $\langle\eta(t)\eta(t')\rangle=2D\delta(t-t')$ with 
$D=k_BT/\gamma=(D_{\parallel}+2D_{\perp})/3$ and 
$\langle\xi(t)\xi(t')\rangle=2D_r\delta(t-t')$. We linearize Eqs.~(\ref{eq:BDsimple1}) 
and (\ref{eq:BDsimple2}) around the stable point $(z^*,\theta^*)=(a/2,\theta_0)$ of 
a fully absorbed particle, and define small perturbation as 
$(\delta z,\delta\theta)=(z-z^*,\theta-\theta^*)$. The equations of motion then reduce 
to an Ornstein-Uhlenbeck process,
\begin{align}
\label{eq:OUprocess2D}
\frac{\mathrm d}{\mathrm d t}
\begin{pmatrix}
\delta z  \\
\delta\theta
\end{pmatrix}
\approx
\begin{pmatrix}
\kappa V\theta_0 & V \\
0 & \frac{\kappa l^2\gamma V\theta_0}{4\gamma_r}
\end{pmatrix}
\begin{pmatrix}
\delta z  \\
\delta\theta
\end{pmatrix}
+
\begin{pmatrix}
\eta  \\
\xi
\end{pmatrix},
\end{align}
where we assume small $\theta_0$. Our aim is to estimate the mean escape time $\tau_e$, 
i.e., the mean time to reach an orientation parallel to the wall, $(z,\theta)=(a_2,0)$, 
from the stable position $(z,\theta)=(z^*,\theta^*)$ by rotational diffusion. In order 
to do so, we reduce the complexity of the problem further by neglecting the motion 
normal to the surface, 
\begin{equation}
\dot{\delta\theta}=-\frac{\mathrm d U}{\mathrm d\delta\theta}+\xi,\quad U=\frac{k}{2}\delta\theta^2.
\label{eq:OUprocess1D}
\end{equation}
Here, $U$ is an effective harmonic potential for the orientation angle with spring 
constant $k=-\kappa l^2\gamma V\theta_0/(4\gamma_r)$. 

The escape problem, Eq.~(\ref{eq:OUprocess1D}), is related to a first passage problem. An 
exact expression of the mean first-passage time from any point along any potential 
to any other point exists \cite{haenggi1990rmp}; however, due to the 
complexity of this expression, an extraction of the leading contributions seems unfeasible.
A low-noise approximation is the Kramers rate theory of crossing a smooth potential 
barrier \cite{haenggi1990rmp,schaar2014arXiv}; here, we use a heuristic 
expression for the mean escape time over a barrier $\Delta U=U(\delta\theta_e)=U(-\theta_0)$, 
which captures the low- and the high-noise limits \cite{drescher2011pnas},
\begin{equation}
\tau_e\approx\frac{\delta\theta_e^2}{D_r}\exp{\left(\frac{\Delta U}{D_r}\right)}=
\frac{\theta_0^2}{D_r}\exp{\left(-\frac{\kappa l^2}{8D}V\theta_0^3\right)}.
\label{eq:kramer}
\end{equation}
Note that $\theta_0<0$ ($\theta_0>0$) for polar (antipolar) particles.

In the simulations, we measure the mean trapping time $\tau_w$ as the time during 
which the swimmer remains within the wall region (range of steric interactions: 
$z\in[0,\delta]$ or $z\in[L-\delta,L]$ with $\delta=l/2+a$). In case of polar swimmers, 
this trapping time is an estimate of $\tau_e$. As can be seen in 
Fig.~\ref{f:t_wall_t_bulk}, we do indeed observe exponential dependencies of $\tau_w$ 
in the low-noise regime, with $\log{(\tau_w)}\propto\theta_0^2$ in 
Fig.~\ref{f:t_wall_t_bulk}(a,b) and $\log{(\tau_w)}\propto V$ in Fig.~\ref{f:t_wall_t_bulk}(c). 
The dependence of $\tau_w$ on $V$ is nicely consistent with Eq.~(\ref{eq:kramer}); 
however, we observe $\log{(\tau_w)}\propto\theta_0^2$ instead of 
$\log{(\tau_e)}\propto\theta_0^3$, as predicted by Eq.~(\ref{eq:kramer}). 
Considering the various approximations in the derivation of Eq.~(\ref{eq:kramer}), 
like the harmonic form of $U(\delta\theta)$ and the dimension reduction, it is not 
surprising that we do not obtain a perfect agreement. In particular, the parabolic 
description of $U$ breaks down with increasing $\delta\theta$; moreover, $U$ is a 
function of $\delta z$ and should soften with increasing $\delta z$.

In case of apolar and antipolar swimmers, which do not point persistently toward the 
wall, the above treatment does not apply. Instead, $\tau_w$ can be deduced from an 
analogy to a semi-flexible polymer adsorbed on a wall \cite{elgeti2009epl,li2009prl}, 
which predicts a scaling behavior $\tau_w\propto V^{-2/3}$, see 
Fig.~\ref{f:t_wall_t_bulk}(c). This process contributes also to $\tau_w$ of polar 
particles, but it becomes negligible as compared to the escape time 
over the effective potential barrier $\Delta U$ for large polarity and $Pe\gg1$. 

The mean time $\tau_b$ a swimmer remains within the bulk region (outside the range of 
steric interactions) is independent of $\theta_0$. 
As can be seen from Fig.~\ref{f:rho_ez_rz}(d), particles leave the wall region 
increasingly parallel with increasing $Pe$. In agreement with the semi-flexible 
polymer analogy \cite{elgeti2009epl}, we observe $\langle\theta\rangle\propto V^{-1/3}$ 
at $z\approx\delta$ in the ballistic regime. Thus, with 
$\tau_b\propto L/(V\sin{\langle\theta\rangle})$ 
we obtain $\tau_b\propto L/V^{2/3}$ consistently with simulations, see 
Fig.~\ref{f:t_wall_t_bulk}(d).

\begin{figure}[t]
\centering
\includegraphics[width=1\columnwidth]{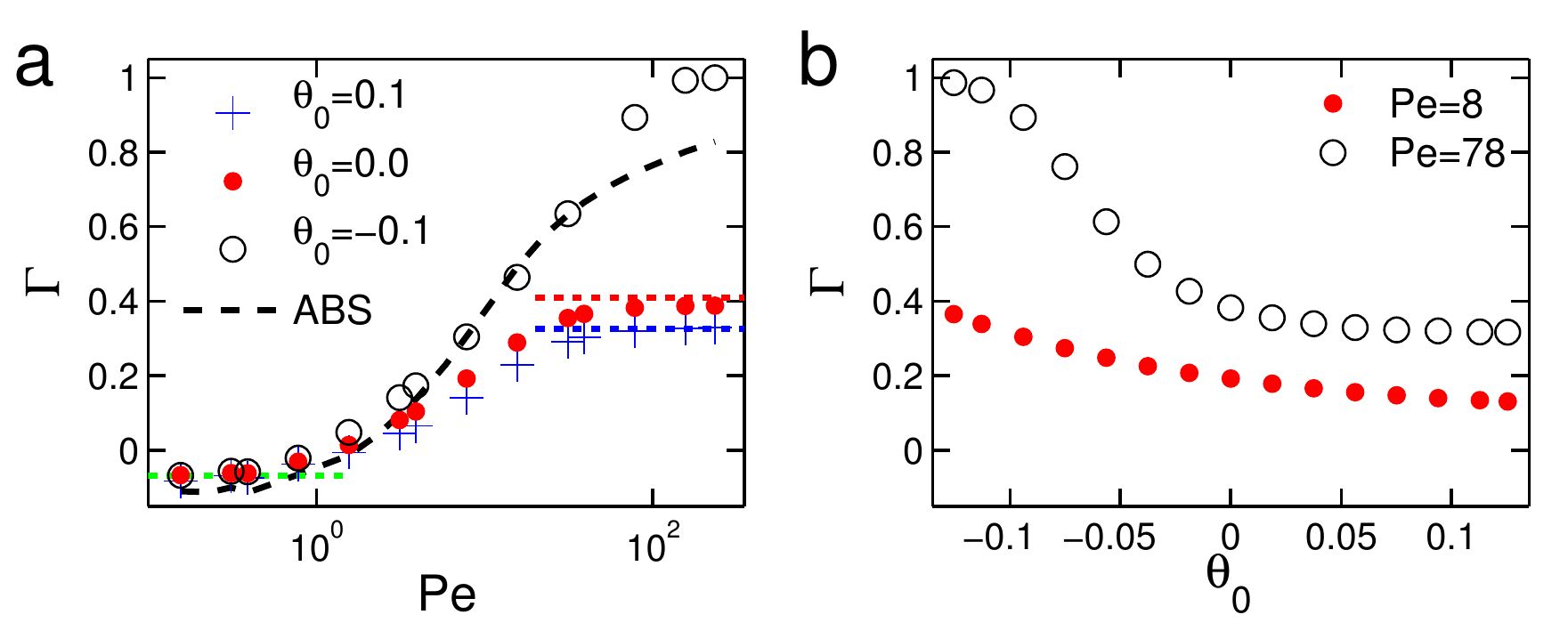}
\caption{\label{f:gamma}(color online) (a) Adsorption $\Gamma$ versus activity $Pe$ 
for antipolar ($\theta_0>0$), apolar ($\theta_0=0$) and polar swimmer ($\theta_0<0$). 
Asymptotic estimates of $\Gamma$ are indicated by dashed lines. $\Gamma$ of active 
Brownian spheres (ABS) is shown for comparison. Note that polar swimmers 
are completely absorbed ($\Gamma=1$) above a critical $Pe$, by contrast, 
apolar and antipolar swimmers are always partially absorbed. 
(b) $\Gamma$ as function of the asymmetry $\theta_0$ at various $Pe$.}
\end{figure}

With this knowledge, it is easy to interpret a global measure of the density distribution, 
the surface excess (or adsorption) $\Gamma=\int_{0}^{L}\left[\rho(z)-\rho_b\right]\,\mathrm{d}z$, 
where $\rho_b$ is the bulk density. 
For a passive hard dumbbell, $\Gamma=-\delta/(L-\delta)<0$, while for fully absorbed particles 
it is $\Gamma=1$. A rough estimate of $\Gamma$ is
\begin{equation}
\Gamma\approx2\int_{0}^{\delta}\rho(z)\,\mathrm{d}z-2\rho_b\delta\approx
\frac{2\tau_w-\tau_b2\delta/(L-2\delta)}{2\tau_w+\tau_b}.
\label{eq:adsorption2}
\end{equation}
Using the scaling of $\tau_w$ and $\tau_b$, see Fig.~\ref{f:t_wall_t_bulk}, 
we obtain low- and high-$Pe$ limits, which are consistent with the simulation results,
as indicated in Fig.~\ref{f:gamma}(a).

\begin{figure}[t]
\centering
\includegraphics[width=0.95\columnwidth]{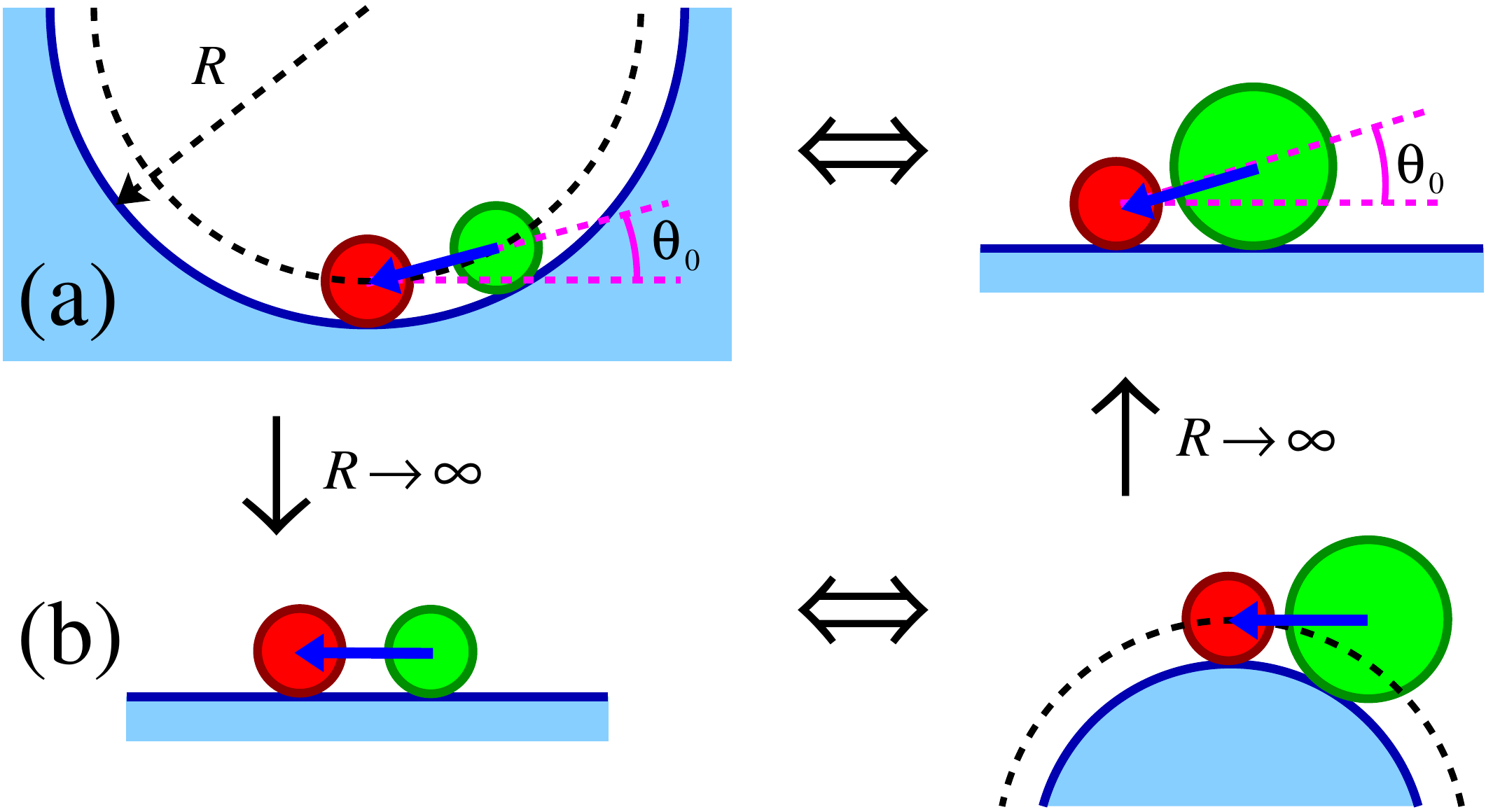}
\caption{\label{f:sketch2}(color online) (a) An apolar swimmer confined 
within a spherical cavity of radius $R$ is equivalent to a polar swimmer 
close to a flat wall provided that the angle $\theta_0\approx-l/(2R)$, 
between the propulsion force of the apolar swimmer and the tangent plane of the cavity 
at the front bead, equals the asymmetry $\theta_0\approx(a_1-a_2)/l$ of the polar particle. 
(b) A polar particle near a convex boundary behaves like an apolar swimmer 
close to a flat wall if $(a_1-a_2)/l\approx-l/(2R)$.}
\end{figure}

The problem of a microswimmer with shape asymmetry moving near a planar wall 
bears a strong resemblance with a swimmer moving near a curved wall, see 
Fig.~\ref{f:sketch2}. Let us consider first a apolar swimmer in spherical 
confinement of radius $R$, see Fig.~\ref{f:sketch2}(a). In this case, as in the case 
of a polar microswimmer at a planar wall, the velocity vector in the stable 
conformation forms an angle with the tangent plane to the wall at the front bead. 
Thus, in both cases, the microswimmer points toward the wall and 
thus should have very long retention times. Secondly, the force of 
a polar microswimmer towards the wall can be partially or fully compensated 
by a convex wall, i.e., for a microswimmer moving 
at the outer surface of a sphere of radius $R$, see Fig.~\ref{f:sketch2}(b). 
In the case of a full compensation, we predict the same accumulation behavior 
as for an apolar particle at a planar wall. Note that an apolar microswimmer 
would strongly scatter at convex wall. Thus, shape polarity provides
the possibility for microswimmers to move along curved surfaces!

Hence, it is obvious to define a generalized asymmetry, considering shape 
asymmetry and wall curvature at once, as $\Theta_0\equiv\theta_0^s+\theta_0^w=(a_1-a_2)/l+kl/(2R)$ 
for $R\gg l\gg|a_1-a_2|$, where $k=+1$ for convex and $k=-1$ for concave boundaries. 
This allows a unified description of asymmetric microswimmers near curved walls,
where $\Theta_0<0$ ($\Theta_0>0$) implies an exponential grow (algebraic decay) 
of $\tau_w$ with $V$. 

\begin{figure}[t]
\centering
\includegraphics[width=\columnwidth]{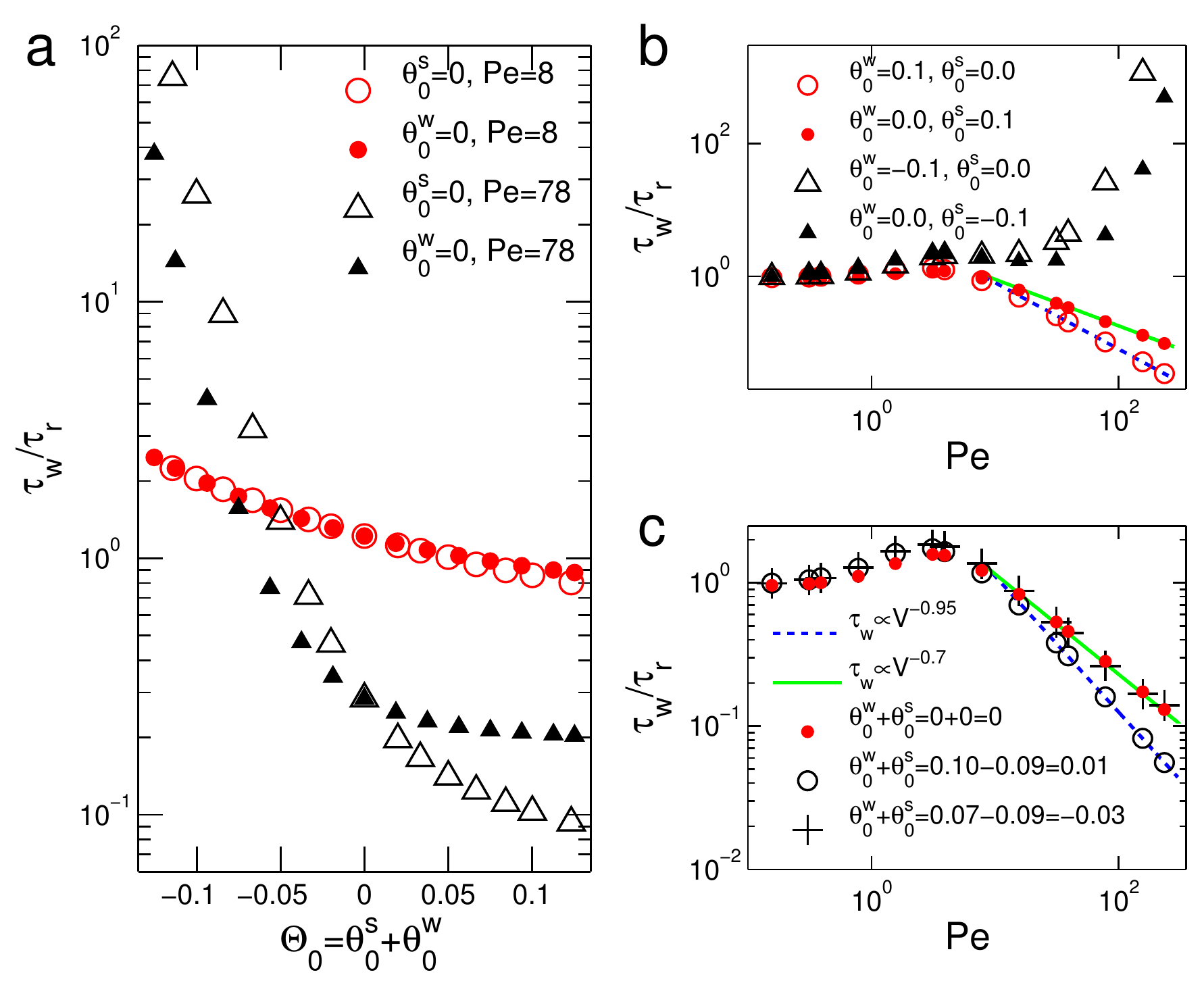}
\caption{\label{f:t_wall_shape_curvature}(color online) Equivalence 
of body shape and wall curvature. $\theta_0^w<0$ corresponds to a motion 
within a spherical cavity and $\theta_0^w>0$ means swimming close to the 
outer surface of a sphere. $\tau_w$ is shown as a function of the generalized 
asymmetry $\Theta_0\equiv\theta_0^s+\theta_0^w$ and $Pe$. 
(a,b) An apolar swimmer ($\theta_0^s=0$) near curved boundaries ($\theta_0^w\neq0$); 
the situation is illustrated in Fig.~\ref{f:sketch2}(a). 
Filled symbols indicate $\tau_w$ of an asymmetric swimmer ($\theta_0^s\neq0$) 
near a flat wall ($\theta_0^w=0$). 
(c) A polar swimmer ($\theta_0^s=-0.09$) near a convex wall ($\theta_0^w>0$); 
the situation is illustrated in Fig.~\ref{f:sketch2}(b). 
In (b) and (c) the same power-law decays are indicated by lines.}
\end{figure}

We have performed various test in order to verify the equivalence 
of shape asymmetry and surface curvature 
\footnote{For the simulations of a microswimmer near a convex wall, we fill 
the space between two spheres of the dumbbell with additional spherical 
segments with diameters interpolating linearly between the front and back spheres, 
similar to Ref.~\cite{wensink2014pre}.
}. 
First, we analyse the behavior of 
an apolar microswimmer ($\theta_0^s=0$) close to a surface with a curvature 
ranging from that of a convex to a concave wall. The wall retention times 
$\tau_w$ as a function of the generalized asymmetry 
$\Theta_0\equiv\theta_0^s+\theta_0^w$ and $Pe$ are 
shown in Fig.~\ref{f:t_wall_shape_curvature}(a,b). The results closely 
resemble the corresponding dependencies of a asymmetric swimmer near a flat wall 
in Fig.~\ref{f:t_wall_t_bulk}(a,b). However, $\tau_w\propto V^{-1}$ for 
$\theta_0^w>0$ in contrast to $\tau_w\propto V^{-0.7}$ for $\theta_0^w=0$; 
the $\tau_w\propto V^{-1}$ behavior reflects a simple ballistic escape 
from convex walls. Further, we have tested whether shape polarity can be 
compensated by a negative curvature of the wall. We simulate a polar swimmer 
with fixed $\theta_0^s=-0.09$ close to the surface of a sphere at different 
radii $R$, i.e., at different wall polarities $\theta_0^{wall}=l/(2R)$. 
The results for $\tau_w$ as a function of $\Theta_0$ and $Pe$ are shown in 
Fig.~\ref{f:t_wall_shape_curvature}(c). Again the similarity with 
the results in Fig.~\ref{f:t_wall_t_bulk}(b) is striking. 
Hence, a symmetric rod near a planar wall is equivalent to a polar 
swimmer near a convex boundary provided that shape and wall polarity 
cancel, i.e., $\Theta_0\approx0\Rightarrow(a_1-a_2)/l\approx-l/(2R)$.

\begin{figure}[t]
\centering
\includegraphics[width=.65\columnwidth]{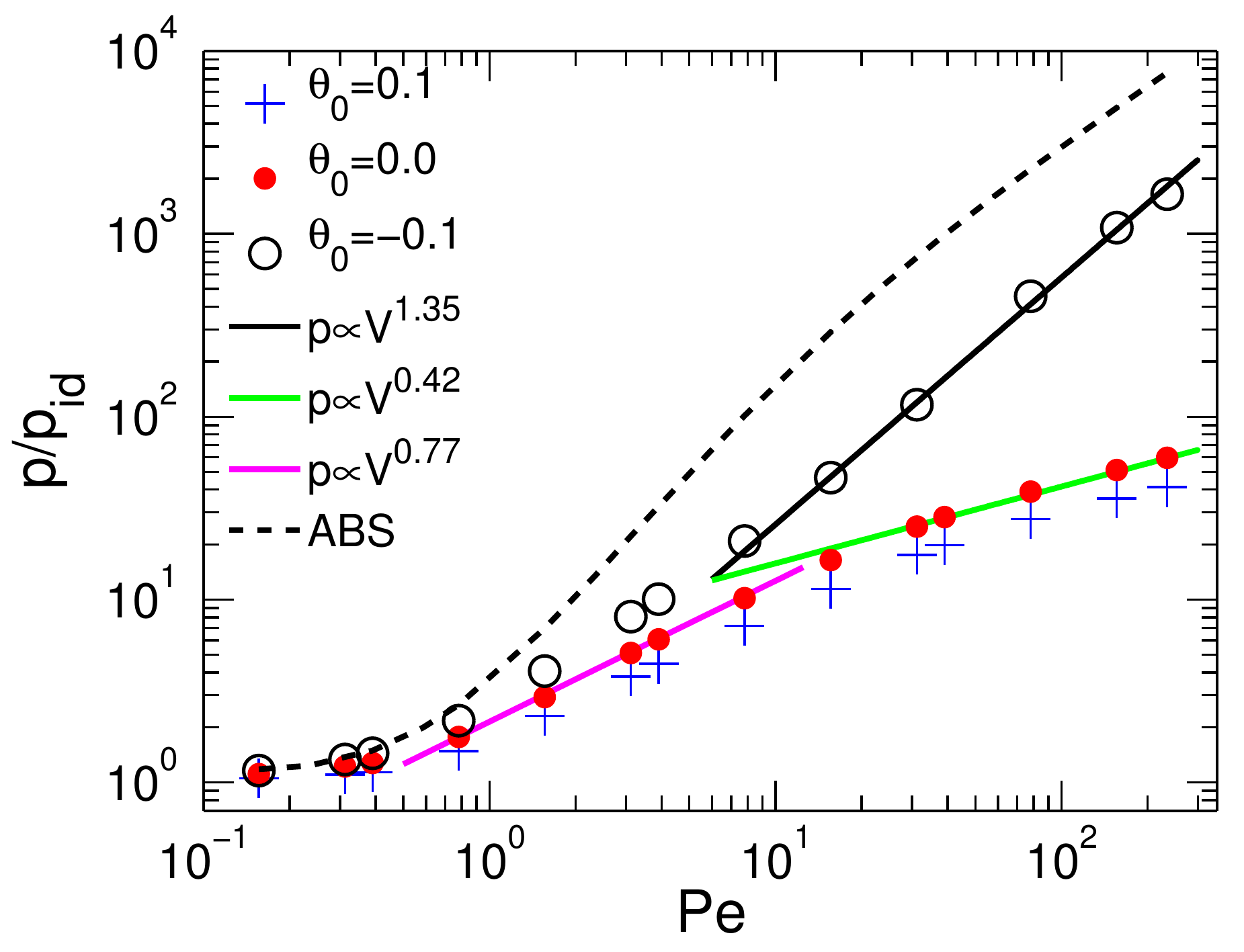}
\caption{\label{f:pressure}(color online) Active pressure $p$ 
as a function of $Pe$ normalized with the ideal gas pressure $p_{id}$. 
At large activity pressure grows superlinear 
for polar swimmers and sublinear for symmetric rods and antipolar particles. 
$p$ of active Brownian spheres (ABS) is shown for comparison \cite{yang2014sm,mallory2014pre}.} 
\end{figure}

Finally, we want to briefly discuss the the wall pressure in active systems.
There are several attempts to construct an equation of state for active fluids 
\cite{yang2014sm,takatori2014prl,ginot2015prx}. However, this idea has been questioned,  
because the pressure $p$, measured as the force exerted on the boundary per wall 
area, should strongly depends on the details of the swimmer-wall interaction 
\cite{solon2014pressure}, in contrast to thermal equilibrium. 
Our results support the latter claim. 
We observe that, in contrast to active Brownian spheres 
where $p\propto V$ in the strong confinement limit ($Pe\gg L/l$) \cite{yang2014sm,mallory2014pre}, 
$p\propto V^{1.35}$ for polar swimmers, and $p\propto V^{0.42}$ for symmetric rods 
and antipolar particles, see Fig.~\ref{f:pressure}. The difference between a spherical and 
a rod-like particle is that the latter exert a significant force only during the 
arrival at the wall, and align immediately parallel with the boundary due to the 
steric torque $T_w(e_z,z)$.

In summary, we have shown that a small shape polarity of microswimmers leads to 
extremly long wall-trapping times.
The exponential dependence of $\tau_w$ on $V$ and $\theta_0$ is responsible for a nearly
complete adsorption of polar particles, in contrast to symmetric rods or antipolar 
particles. The pressure $p$ in active system is sensitive to the details of the swimmer-wall 
interaction \cite{solon2014pressure}, in particular the variation of the asymmetry 
from antipolar to polar change the grow of $p$ with $V$ from sublinear to superlinear. 

We have neglected HI in our analysis to elucidate the effects of swimmer shape 
on wall adsorption. However, it is worthwhile to briefly speculate on the interplay of
both effects. We restrict our discussion to pushers, i.e., to micro-organisms and 
self-propelled particles that generate thrust behind the body, such as sperm or bacteria. 
The retention time due to HI, $\tau_e^{HI}$, corresponds to the time to reach an 
angle $\theta_e^{HI}>0$ starting from parallel, where self-propulsion outweighs  
hydrodynamic attraction \cite{drescher2011pnas}. $\tau_e^{HI}$ has been predicted to follow 
an Arrhenius-Kramer-like behavior \cite{drescher2011pnas,schaar2014arXiv}. 
For a polar pusher, the effective potential for the orientation angle now
contains two barriers at $\theta=0$ (due to shape) and $\theta=\theta_e^{HI}$ 
(due to HI). Thus, a naive expectation would be that the resulting retention time 
is a sum of the retention times due to shape and HI \cite{risken1984fokker}. 
However, we expect a synergistic effect, because hydrodynamic interaction 
will both additionally attract the swimmer towards the wall and further restrict 
rotational movement, which increases the effective barrier height.
In contrast, for antipolar pushers, hydrodynamic attraction will favor 
a configuration with orientation pointing away from the wall, and hence can be
expected to shorten the way to the potential barrier due to HI at $\theta_e^{HI}>0$,
thereby ultimately decreasing HI-induced wall adhesion.

Furthermore, we have shown that wall curvature can compensate the effect 
of shape asymmetry, i.e., asymmetry and curvature are two sides of the same coin. 
Thus, the combination of both effects can be used in order 
to design artificial microswimmers or microfluidic devices for particular tasks. 
For example, microswimmers could be designed which move along surfaces 
within a porous medium \cite{takagi2014sm,brown2014swimming}, 
while corrugated microfluidic channels can be constructed to reduce wall 
accumulation \cite{denissenko2012pnas,guidobaldi2015bmf}.

We thank J. Hu, G. Sch\"utz, A. Varghese, and R. G. Winkler 
for helpful discussions. We gratefully acknowledge support by the DFG within 
the priority program SPP 1726 ``Microswimmers".


\begin{thebibliography}{36}%
\makeatletter
\providecommand \@ifxundefined [1]{%
 \@ifx{#1\undefined}
}%
\providecommand \@ifnum [1]{%
 \ifnum #1\expandafter \@firstoftwo
 \else \expandafter \@secondoftwo
 \fi
}%
\providecommand \@ifx [1]{%
 \ifx #1\expandafter \@firstoftwo
 \else \expandafter \@secondoftwo
 \fi
}%
\providecommand \natexlab [1]{#1}%
\providecommand \enquote  [1]{``#1''}%
\providecommand \bibnamefont  [1]{#1}%
\providecommand \bibfnamefont [1]{#1}%
\providecommand \citenamefont [1]{#1}%
\providecommand \href@noop [0]{\@secondoftwo}%
\providecommand \href [0]{\begingroup \@sanitize@url \@href}%
\providecommand \@href[1]{\@@startlink{#1}\@@href}%
\providecommand \@@href[1]{\endgroup#1\@@endlink}%
\providecommand \@sanitize@url [0]{\catcode `\\12\catcode `\$12\catcode
  `\&12\catcode `\#12\catcode `\^12\catcode `\_12\catcode `\%12\relax}%
\providecommand \@@startlink[1]{}%
\providecommand \@@endlink[0]{}%
\providecommand \url  [0]{\begingroup\@sanitize@url \@url }%
\providecommand \@url [1]{\endgroup\@href {#1}{\urlprefix }}%
\providecommand \urlprefix  [0]{URL }%
\providecommand \Eprint [0]{\href }%
\providecommand \doibase [0]{http://dx.doi.org/}%
\providecommand \selectlanguage [0]{\@gobble}%
\providecommand \bibinfo  [0]{\@secondoftwo}%
\providecommand \bibfield  [0]{\@secondoftwo}%
\providecommand \translation [1]{[#1]}%
\providecommand \BibitemOpen [0]{}%
\providecommand \bibitemStop [0]{}%
\providecommand \bibitemNoStop [0]{.\EOS\space}%
\providecommand \EOS [0]{\spacefactor3000\relax}%
\providecommand \BibitemShut  [1]{\csname bibitem#1\endcsname}%
\let\auto@bib@innerbib\@empty
\bibitem [{\citenamefont {Lauga}\ and\ \citenamefont
  {Powers}(2009)}]{lauga2009rpp}%
  \BibitemOpen
  \bibfield  {author} {\bibinfo {author} {\bibfnamefont {E.}~\bibnamefont
  {Lauga}}\ and\ \bibinfo {author} {\bibfnamefont {T.~R.}\ \bibnamefont
  {Powers}},\ }\href@noop {} {\bibfield  {journal} {\bibinfo  {journal}
  {Reports on Progress in Physics}\ }\textbf {\bibinfo {volume} {72}},\
  \bibinfo {pages} {096601} (\bibinfo {year} {2009})}\BibitemShut {NoStop}%
\bibitem [{\citenamefont {Elgeti}\ \emph {et~al.}(2015)\citenamefont {Elgeti},
  \citenamefont {Winkler},\ and\ \citenamefont {Gompper}}]{elgeti2015rpp}%
  \BibitemOpen
  \bibfield  {author} {\bibinfo {author} {\bibfnamefont {J.}~\bibnamefont
  {Elgeti}}, \bibinfo {author} {\bibfnamefont {R.~G.}\ \bibnamefont {Winkler}},
  \ and\ \bibinfo {author} {\bibfnamefont {G.}~\bibnamefont {Gompper}},\ }\href
  {\doibase 10.1088/0034-4885/78/5/056601} {\bibfield  {journal} {\bibinfo
  {journal} {Rep. Prog. Phys.}\ }\textbf {\bibinfo {volume} {78}},\ \bibinfo
  {pages} {056601} (\bibinfo {year} {2015})}\BibitemShut {NoStop}%
\bibitem [{\citenamefont {Tuson}\ and\ \citenamefont
  {Weibel}(2013)}]{tuson2013sm}%
  \BibitemOpen
  \bibfield  {author} {\bibinfo {author} {\bibfnamefont {H.~H.}\ \bibnamefont
  {Tuson}}\ and\ \bibinfo {author} {\bibfnamefont {D.~B.}\ \bibnamefont
  {Weibel}},\ }\href {\doibase 10.1039/C3SM27705D} {\bibfield  {journal}
  {\bibinfo  {journal} {Soft Matter}\ }\textbf {\bibinfo {volume} {9}},\
  \bibinfo {pages} {4368} (\bibinfo {year} {2013})}\BibitemShut {NoStop}%
\bibitem [{\citenamefont {Berke}\ \emph {et~al.}(2008)\citenamefont {Berke},
  \citenamefont {Turner}, \citenamefont {Berg},\ and\ \citenamefont
  {Lauga}}]{berke2008prl}%
  \BibitemOpen
  \bibfield  {author} {\bibinfo {author} {\bibfnamefont {A.~P.}\ \bibnamefont
  {Berke}}, \bibinfo {author} {\bibfnamefont {L.}~\bibnamefont {Turner}},
  \bibinfo {author} {\bibfnamefont {H.~C.}\ \bibnamefont {Berg}}, \ and\
  \bibinfo {author} {\bibfnamefont {E.}~\bibnamefont {Lauga}},\ }\href
  {\doibase 10.1103/PhysRevLett.101.038102} {\bibfield  {journal} {\bibinfo
  {journal} {Phys.~Rev.~Lett.}\ }\textbf {\bibinfo {volume} {101}},\ \bibinfo
  {pages} {038102} (\bibinfo {year} {2008})}\BibitemShut {NoStop}%
\bibitem [{\citenamefont {Drescher}\ \emph {et~al.}(2011)\citenamefont
  {Drescher}, \citenamefont {Dunkel}, \citenamefont {Cisneros}, \citenamefont
  {Ganguly},\ and\ \citenamefont {Goldstein}}]{drescher2011pnas}%
  \BibitemOpen
  \bibfield  {author} {\bibinfo {author} {\bibfnamefont {K.}~\bibnamefont
  {Drescher}}, \bibinfo {author} {\bibfnamefont {J.}~\bibnamefont {Dunkel}},
  \bibinfo {author} {\bibfnamefont {L.~H.}\ \bibnamefont {Cisneros}}, \bibinfo
  {author} {\bibfnamefont {S.}~\bibnamefont {Ganguly}}, \ and\ \bibinfo
  {author} {\bibfnamefont {R.~E.}\ \bibnamefont {Goldstein}},\ }\href {\doibase
  10.1073/pnas.1019079108} {\bibfield  {journal} {\bibinfo  {journal}
  {Proc.~Natl.~Acad.~Sci.~USA}\ }\textbf {\bibinfo {volume} {108}},\ \bibinfo
  {pages} {10940} (\bibinfo {year} {2011})}\BibitemShut {NoStop}%
\bibitem [{\citenamefont {Elgeti}\ and\ \citenamefont
  {Gompper}(2009)}]{elgeti2009epl}%
  \BibitemOpen
  \bibfield  {author} {\bibinfo {author} {\bibfnamefont {J.}~\bibnamefont
  {Elgeti}}\ and\ \bibinfo {author} {\bibfnamefont {G.}~\bibnamefont
  {Gompper}},\ }\href@noop {} {\bibfield  {journal} {\bibinfo  {journal} {EPL}\
  }\textbf {\bibinfo {volume} {85}},\ \bibinfo {pages} {38002} (\bibinfo {year}
  {2009})}\BibitemShut {NoStop}%
\bibitem [{\citenamefont {Li}\ and\ \citenamefont {Tang}(2009)}]{li2009prl}%
  \BibitemOpen
  \bibfield  {author} {\bibinfo {author} {\bibfnamefont {G.}~\bibnamefont
  {Li}}\ and\ \bibinfo {author} {\bibfnamefont {J.~X.}\ \bibnamefont {Tang}},\
  }\href {\doibase 10.1103/PhysRevLett.103.078101} {\bibfield  {journal}
  {\bibinfo  {journal} {Phys.~Rev.~Lett.}\ }\textbf {\bibinfo {volume} {103}},\
  \bibinfo {pages} {078101} (\bibinfo {year} {2009})}\BibitemShut {NoStop}%
\bibitem [{\citenamefont {Kantsler}\ \emph {et~al.}(2013)\citenamefont
  {Kantsler}, \citenamefont {Dunkel}, \citenamefont {Polin},\ and\
  \citenamefont {Goldstein}}]{kantsler2013pnas}%
  \BibitemOpen
  \bibfield  {author} {\bibinfo {author} {\bibfnamefont {V.}~\bibnamefont
  {Kantsler}}, \bibinfo {author} {\bibfnamefont {J.}~\bibnamefont {Dunkel}},
  \bibinfo {author} {\bibfnamefont {M.}~\bibnamefont {Polin}}, \ and\ \bibinfo
  {author} {\bibfnamefont {R.~E.}\ \bibnamefont {Goldstein}},\ }\href {\doibase
  10.1073/pnas.1210548110} {\bibfield  {journal} {\bibinfo  {journal}
  {Proc.~Natl.~Acad.~Sci.~USA}\ }\textbf {\bibinfo {volume} {110}},\ \bibinfo
  {pages} {1187} (\bibinfo {year} {2013})}\BibitemShut {NoStop}%
\bibitem [{\citenamefont {Denissenko}\ \emph {et~al.}(2012)\citenamefont
  {Denissenko}, \citenamefont {Kantsler}, \citenamefont {Smith},\ and\
  \citenamefont {Kirkman-Brown}}]{denissenko2012pnas}%
  \BibitemOpen
  \bibfield  {author} {\bibinfo {author} {\bibfnamefont {P.}~\bibnamefont
  {Denissenko}}, \bibinfo {author} {\bibfnamefont {V.}~\bibnamefont
  {Kantsler}}, \bibinfo {author} {\bibfnamefont {D.~J.}\ \bibnamefont {Smith}},
  \ and\ \bibinfo {author} {\bibfnamefont {J.}~\bibnamefont {Kirkman-Brown}},\
  }\href {\doibase 10.1073/pnas.1202934109} {\bibfield  {journal} {\bibinfo
  {journal} {Proc.~Natl.~Acad.~Sci.~USA}\ }\textbf {\bibinfo {volume} {109}},\
  \bibinfo {pages} {8007} (\bibinfo {year} {2012})}\BibitemShut {NoStop}%
\bibitem [{\citenamefont {Elgeti}\ \emph {et~al.}(2010)\citenamefont {Elgeti},
  \citenamefont {Kaupp},\ and\ \citenamefont {Gompper}}]{elgeti10bj}%
  \BibitemOpen
  \bibfield  {author} {\bibinfo {author} {\bibfnamefont {J.}~\bibnamefont
  {Elgeti}}, \bibinfo {author} {\bibfnamefont {U.~B.}\ \bibnamefont {Kaupp}}, \
  and\ \bibinfo {author} {\bibfnamefont {G.}~\bibnamefont {Gompper}},\ }\href
  {\doibase 10.1016/j.bpj.2010.05.015} {\bibfield  {journal} {\bibinfo
  {journal} {Biophys.~J.}\ }\textbf {\bibinfo {volume} {99}},\ \bibinfo {pages}
  {1018} (\bibinfo {year} {2010})}\BibitemShut {NoStop}%
\bibitem [{\citenamefont {Elgeti}\ and\ \citenamefont
  {Gompper}(2013)}]{elgeti2013epl}%
  \BibitemOpen
  \bibfield  {author} {\bibinfo {author} {\bibfnamefont {J.}~\bibnamefont
  {Elgeti}}\ and\ \bibinfo {author} {\bibfnamefont {G.}~\bibnamefont
  {Gompper}},\ }\href@noop {} {\bibfield  {journal} {\bibinfo  {journal} {EPL}\
  }\textbf {\bibinfo {volume} {101}},\ \bibinfo {pages} {48003} (\bibinfo
  {year} {2013})}\BibitemShut {NoStop}%
\bibitem [{\citenamefont {Lee}(2013)}]{lee2013njp}%
  \BibitemOpen
  \bibfield  {author} {\bibinfo {author} {\bibfnamefont {C.~F.}\ \bibnamefont
  {Lee}},\ }\href@noop {} {\bibfield  {journal} {\bibinfo  {journal}
  {New~J.~Phys.}\ }\textbf {\bibinfo {volume} {15}},\ \bibinfo {pages} {055007}
  (\bibinfo {year} {2013})}\BibitemShut {NoStop}%
\bibitem [{\citenamefont {Wensink}\ \emph {et~al.}(2014)\citenamefont
  {Wensink}, \citenamefont {Kantsler}, \citenamefont {Goldstein},\ and\
  \citenamefont {Dunkel}}]{wensink2014pre}%
  \BibitemOpen
  \bibfield  {author} {\bibinfo {author} {\bibfnamefont {H.~H.}\ \bibnamefont
  {Wensink}}, \bibinfo {author} {\bibfnamefont {V.}~\bibnamefont {Kantsler}},
  \bibinfo {author} {\bibfnamefont {R.~E.}\ \bibnamefont {Goldstein}}, \ and\
  \bibinfo {author} {\bibfnamefont {J.}~\bibnamefont {Dunkel}},\ }\href
  {\doibase 10.1103/PhysRevE.89.010302} {\bibfield  {journal} {\bibinfo
  {journal} {Phys.~Rev.~E}\ }\textbf {\bibinfo {volume} {89}},\ \bibinfo
  {pages} {010302} (\bibinfo {year} {2014})}\BibitemShut {NoStop}%
\bibitem [{\citenamefont {Spagnolie}\ \emph {et~al.}(2015)\citenamefont
  {Spagnolie}, \citenamefont {Moreno-Flores}, \citenamefont {Bartolo},\ and\
  \citenamefont {Lauga}}]{spagnolie2014sm}%
  \BibitemOpen
  \bibfield  {author} {\bibinfo {author} {\bibfnamefont {S.~E.}\ \bibnamefont
  {Spagnolie}}, \bibinfo {author} {\bibfnamefont {G.~R.}\ \bibnamefont
  {Moreno-Flores}}, \bibinfo {author} {\bibfnamefont {D.}~\bibnamefont
  {Bartolo}}, \ and\ \bibinfo {author} {\bibfnamefont {E.}~\bibnamefont
  {Lauga}},\ }\href {\doibase 10.1039/C4SM02785J} {\bibfield  {journal}
  {\bibinfo  {journal} {Soft Matter}\ }\textbf {\bibinfo {volume} {11}},\
  \bibinfo {pages} {3396} (\bibinfo {year} {2015})}\BibitemShut {NoStop}%
\bibitem [{\citenamefont {Schaar}\ \emph {et~al.}(2014)\citenamefont {Schaar},
  \citenamefont {Z{\"o}ttl},\ and\ \citenamefont {Stark}}]{schaar2014arXiv}%
  \BibitemOpen
  \bibfield  {author} {\bibinfo {author} {\bibfnamefont {K.}~\bibnamefont
  {Schaar}}, \bibinfo {author} {\bibfnamefont {A.}~\bibnamefont {Z{\"o}ttl}}, \
  and\ \bibinfo {author} {\bibfnamefont {H.}~\bibnamefont {Stark}},\
  }\href@noop {} {\bibfield  {journal} {\bibinfo  {journal} {arXiv:1412.6435}\
  } (\bibinfo {year} {2014})}\BibitemShut {NoStop}%
\bibitem [{\citenamefont {Kaehr}\ and\ \citenamefont
  {Shear}(2009)}]{kaehr2009lc}%
  \BibitemOpen
  \bibfield  {author} {\bibinfo {author} {\bibfnamefont {B.}~\bibnamefont
  {Kaehr}}\ and\ \bibinfo {author} {\bibfnamefont {J.~B.}\ \bibnamefont
  {Shear}},\ }\href {\doibase 10.1039/B908119D} {\bibfield  {journal} {\bibinfo
   {journal} {Lab Chip}\ }\textbf {\bibinfo {volume} {9}},\ \bibinfo {pages}
  {2632} (\bibinfo {year} {2009})}\BibitemShut {NoStop}%
\bibitem [{\citenamefont {Wioland}\ \emph {et~al.}(2013)\citenamefont
  {Wioland}, \citenamefont {Woodhouse}, \citenamefont {Dunkel}, \citenamefont
  {Kessler},\ and\ \citenamefont {Goldstein}}]{wioland2013prl}%
  \BibitemOpen
  \bibfield  {author} {\bibinfo {author} {\bibfnamefont {H.}~\bibnamefont
  {Wioland}}, \bibinfo {author} {\bibfnamefont {F.~G.}\ \bibnamefont
  {Woodhouse}}, \bibinfo {author} {\bibfnamefont {J.}~\bibnamefont {Dunkel}},
  \bibinfo {author} {\bibfnamefont {J.~O.}\ \bibnamefont {Kessler}}, \ and\
  \bibinfo {author} {\bibfnamefont {R.~E.}\ \bibnamefont {Goldstein}},\ }\href
  {\doibase 10.1103/PhysRevLett.110.268102} {\bibfield  {journal} {\bibinfo
  {journal} {Phys.~Rev.~Lett.}\ }\textbf {\bibinfo {volume} {110}},\ \bibinfo
  {pages} {268102} (\bibinfo {year} {2013})}\BibitemShut {NoStop}%
\bibitem [{\citenamefont {van Teeffelen}\ \emph {et~al.}(2009)\citenamefont
  {van Teeffelen}, \citenamefont {Zimmermann},\ and\ \citenamefont
  {L{\"o}wen}}]{teeffelen2009sm}%
  \BibitemOpen
  \bibfield  {author} {\bibinfo {author} {\bibfnamefont {S.}~\bibnamefont {van
  Teeffelen}}, \bibinfo {author} {\bibfnamefont {U.}~\bibnamefont
  {Zimmermann}}, \ and\ \bibinfo {author} {\bibfnamefont {H.}~\bibnamefont
  {L{\"o}wen}},\ }\href {\doibase 10.1039/B911365G} {\bibfield  {journal}
  {\bibinfo  {journal} {Soft Matter}\ }\textbf {\bibinfo {volume} {5}},\
  \bibinfo {pages} {4510} (\bibinfo {year} {2009})}\BibitemShut {NoStop}%
\bibitem [{\citenamefont {Fily}\ \emph {et~al.}(2014)\citenamefont {Fily},
  \citenamefont {Baskaran},\ and\ \citenamefont {Hagan}}]{fily2014sm}%
  \BibitemOpen
  \bibfield  {author} {\bibinfo {author} {\bibfnamefont {Y.}~\bibnamefont
  {Fily}}, \bibinfo {author} {\bibfnamefont {A.}~\bibnamefont {Baskaran}}, \
  and\ \bibinfo {author} {\bibfnamefont {M.~F.}\ \bibnamefont {Hagan}},\ }\href
  {\doibase 10.1039/C4SM00975D} {\bibfield  {journal} {\bibinfo  {journal}
  {Soft Matter}\ }\textbf {\bibinfo {volume} {10}},\ \bibinfo {pages} {5609}
  (\bibinfo {year} {2014})}\BibitemShut {NoStop}%
\bibitem [{\citenamefont {Takagi}\ \emph {et~al.}(2014)\citenamefont {Takagi},
  \citenamefont {Palacci}, \citenamefont {Braunschweig}, \citenamefont
  {Shelley},\ and\ \citenamefont {Zhang}}]{takagi2014sm}%
  \BibitemOpen
  \bibfield  {author} {\bibinfo {author} {\bibfnamefont {D.}~\bibnamefont
  {Takagi}}, \bibinfo {author} {\bibfnamefont {J.}~\bibnamefont {Palacci}},
  \bibinfo {author} {\bibfnamefont {A.~B.}\ \bibnamefont {Braunschweig}},
  \bibinfo {author} {\bibfnamefont {M.~J.}\ \bibnamefont {Shelley}}, \ and\
  \bibinfo {author} {\bibfnamefont {J.}~\bibnamefont {Zhang}},\ }\href
  {\doibase 10.1039/C3SM52815D} {\bibfield  {journal} {\bibinfo  {journal}
  {Soft Matter}\ }\textbf {\bibinfo {volume} {10}},\ \bibinfo {pages} {1784}
  (\bibinfo {year} {2014})}\BibitemShut {NoStop}%
\bibitem [{\citenamefont {Vladescu}\ \emph {et~al.}(2014)\citenamefont
  {Vladescu}, \citenamefont {Marsden}, \citenamefont {Schwarz-Linek},
  \citenamefont {Martinez}, \citenamefont {Arlt}, \citenamefont {Morozov},
  \citenamefont {Marenduzzo}, \citenamefont {Cates},\ and\ \citenamefont
  {Poon}}]{vladescu2014prl}%
  \BibitemOpen
  \bibfield  {author} {\bibinfo {author} {\bibfnamefont {I.~D.}\ \bibnamefont
  {Vladescu}}, \bibinfo {author} {\bibfnamefont {E.~J.}\ \bibnamefont
  {Marsden}}, \bibinfo {author} {\bibfnamefont {J.}~\bibnamefont
  {Schwarz-Linek}}, \bibinfo {author} {\bibfnamefont {V.~A.}\ \bibnamefont
  {Martinez}}, \bibinfo {author} {\bibfnamefont {J.}~\bibnamefont {Arlt}},
  \bibinfo {author} {\bibfnamefont {A.~N.}\ \bibnamefont {Morozov}}, \bibinfo
  {author} {\bibfnamefont {D.}~\bibnamefont {Marenduzzo}}, \bibinfo {author}
  {\bibfnamefont {M.~E.}\ \bibnamefont {Cates}}, \ and\ \bibinfo {author}
  {\bibfnamefont {W.~C.~K.}\ \bibnamefont {Poon}},\ }\href {\doibase
  10.1103/PhysRevLett.113.268101} {\bibfield  {journal} {\bibinfo  {journal}
  {Phys.~Rev.~Lett.}\ }\textbf {\bibinfo {volume} {113}},\ \bibinfo {pages}
  {268101} (\bibinfo {year} {2014})}\BibitemShut {NoStop}%
\bibitem [{\citenamefont {Guidobaldi}\ \emph {et~al.}(2015)\citenamefont
  {Guidobaldi}, \citenamefont {Jeyaram}, \citenamefont {Condat}, \citenamefont
  {Oviedo}, \citenamefont {Berdakin}, \citenamefont {Moshchalkov},
  \citenamefont {Giojalas}, \citenamefont {Silhanek},\ and\ \citenamefont
  {Marconi}}]{guidobaldi2015bmf}%
  \BibitemOpen
  \bibfield  {author} {\bibinfo {author} {\bibfnamefont {H.~A.}\ \bibnamefont
  {Guidobaldi}}, \bibinfo {author} {\bibfnamefont {Y.}~\bibnamefont {Jeyaram}},
  \bibinfo {author} {\bibfnamefont {C.~A.}\ \bibnamefont {Condat}}, \bibinfo
  {author} {\bibfnamefont {M.}~\bibnamefont {Oviedo}}, \bibinfo {author}
  {\bibfnamefont {I.}~\bibnamefont {Berdakin}}, \bibinfo {author}
  {\bibfnamefont {V.~V.}\ \bibnamefont {Moshchalkov}}, \bibinfo {author}
  {\bibfnamefont {L.~C.}\ \bibnamefont {Giojalas}}, \bibinfo {author}
  {\bibfnamefont {A.~V.}\ \bibnamefont {Silhanek}}, \ and\ \bibinfo {author}
  {\bibfnamefont {V.~I.}\ \bibnamefont {Marconi}},\ }\href {\doibase
  http://dx.doi.org/10.1063/1.4918979} {\bibfield  {journal} {\bibinfo
  {journal} {Biomicrofluidics}\ }\textbf {\bibinfo {volume} {9}},\ \bibinfo
  {eid} {024122} (\bibinfo {year} {2015})}\BibitemShut {NoStop}%
\bibitem [{Note1()}]{Note1}%
  \BibitemOpen
  \bibinfo {note} {The sphere $\alpha \in \protect \{1,2\protect \}$ interacts
  with the lower wall via \begin {equation} \protect \frac {U_{\alpha
  }}{k_BT}=10\protect \frac {\protect \qopname \relax o{exp}{\left [-\kappa
  (z_{\alpha }-a_{\alpha })\right ]}}{\kappa (z_{\alpha }-a_{\alpha })}, \label
  {eq:yukawa} \end {equation} and equivalently with the upper wall, where
  $z_{\alpha }=\protect \mathbf {r}_{\alpha }\protect \mathaccentV
  {hat}05E{\protect \mathbf {z}}$ is the $z$-coordinate of sphere $\alpha $.
  Strong screening is achieved by using $\kappa a=10$.}\BibitemShut {Stop}%
\bibitem [{\citenamefont {L\"owen}(1994)}]{loewen1994pre}%
  \BibitemOpen
  \bibfield  {author} {\bibinfo {author} {\bibfnamefont {H.}~\bibnamefont
  {L\"owen}},\ }\href {\doibase 10.1103/PhysRevE.50.1232} {\bibfield  {journal}
  {\bibinfo  {journal} {Phys.~Rev.~E}\ }\textbf {\bibinfo {volume} {50}},\
  \bibinfo {pages} {1232} (\bibinfo {year} {1994})}\BibitemShut {NoStop}%
\bibitem [{Note2()}]{Note2}%
  \BibitemOpen
  \bibinfo {note} {Using realistic parameters for {\protect \it Escherichia
  coli} \cite {drescher2011pnas}, we set length scales to $l=8$ $\mu $m and
  $a=a_1+a_2=1$ $\mu $m, diffusion constants to $D_{\parallel }=0.149$ $\mu
  $m$^2/$s, $D_{\perp }=0.135$ $\mu $m$^2/$s, and $D_r=0.032$ s$^{-1}$. We vary
  $Pe$ by changing $V$ up to $60$ $\mu $ms$^{-1}$.}\BibitemShut {Stop}%
\bibitem [{\citenamefont {Tung}\ \emph {et~al.}(2015)\citenamefont {Tung},
  \citenamefont {Ardon}, \citenamefont {Roy}, \citenamefont {Koch},
  \citenamefont {Suarez},\ and\ \citenamefont {Wu}}]{tung2015prl}%
  \BibitemOpen
  \bibfield  {author} {\bibinfo {author} {\bibfnamefont {C.-K.}\ \bibnamefont
  {Tung}}, \bibinfo {author} {\bibfnamefont {F.}~\bibnamefont {Ardon}},
  \bibinfo {author} {\bibfnamefont {A.}~\bibnamefont {Roy}}, \bibinfo {author}
  {\bibfnamefont {D.~L.}\ \bibnamefont {Koch}}, \bibinfo {author}
  {\bibfnamefont {S.~S.}\ \bibnamefont {Suarez}}, \ and\ \bibinfo {author}
  {\bibfnamefont {M.}~\bibnamefont {Wu}},\ }\href {\doibase
  10.1103/PhysRevLett.114.108102} {\bibfield  {journal} {\bibinfo  {journal}
  {Phys. Rev. Lett.}\ }\textbf {\bibinfo {volume} {114}},\ \bibinfo {pages}
  {108102} (\bibinfo {year} {2015})}\BibitemShut {NoStop}%
\bibitem [{\citenamefont {Raible}\ and\ \citenamefont
  {Engel}(2004)}]{raible2004aomc}%
  \BibitemOpen
  \bibfield  {author} {\bibinfo {author} {\bibfnamefont {M.}~\bibnamefont
  {Raible}}\ and\ \bibinfo {author} {\bibfnamefont {A.}~\bibnamefont {Engel}},\
  }\href {\doibase 10.1002/aoc.757} {\bibfield  {journal} {\bibinfo  {journal}
  {Appl.~Organomet.~Chem.}\ }\textbf {\bibinfo {volume} {18}},\ \bibinfo
  {pages} {536} (\bibinfo {year} {2004})}\BibitemShut {NoStop}%
\bibitem [{\citenamefont {H{\"a}nggi}\ \emph {et~al.}(1990)\citenamefont
  {H{\"a}nggi}, \citenamefont {Talkner},\ and\ \citenamefont
  {Borkovec}}]{haenggi1990rmp}%
  \BibitemOpen
  \bibfield  {author} {\bibinfo {author} {\bibfnamefont {P.}~\bibnamefont
  {H{\"a}nggi}}, \bibinfo {author} {\bibfnamefont {P.}~\bibnamefont {Talkner}},
  \ and\ \bibinfo {author} {\bibfnamefont {M.}~\bibnamefont {Borkovec}},\
  }\href {\doibase 10.1103/RevModPhys.62.251} {\bibfield  {journal} {\bibinfo
  {journal} {Rev. Mod. Phys.}\ }\textbf {\bibinfo {volume} {62}},\ \bibinfo
  {pages} {251} (\bibinfo {year} {1990})}\BibitemShut {NoStop}%
\bibitem [{Note3()}]{Note3}%
  \BibitemOpen
  \bibinfo {note} {{\protect For the simulations of a microswimmer
  near a convex wall, we fill the space between two spheres of the dumbbell
  with additional spherical segments with diameters interpolating linearly
  between the front and back spheres, similar to Ref.~\cite
  {wensink2014pre}.}}\BibitemShut {Stop}%
\bibitem [{\citenamefont {Yang}\ \emph {et~al.}(2014)\citenamefont {Yang},
  \citenamefont {Manning},\ and\ \citenamefont {Marchetti}}]{yang2014sm}%
  \BibitemOpen
  \bibfield  {author} {\bibinfo {author} {\bibfnamefont {X.}~\bibnamefont
  {Yang}}, \bibinfo {author} {\bibfnamefont {M.~L.}\ \bibnamefont {Manning}}, \
  and\ \bibinfo {author} {\bibfnamefont {M.~C.}\ \bibnamefont {Marchetti}},\
  }\href {\doibase 10.1039/C4SM00927D} {\bibfield  {journal} {\bibinfo
  {journal} {Soft Matter}\ }\textbf {\bibinfo {volume} {10}},\ \bibinfo {pages}
  {6477} (\bibinfo {year} {2014})}\BibitemShut {NoStop}%
\bibitem [{\citenamefont {Takatori}\ \emph {et~al.}(2014)\citenamefont
  {Takatori}, \citenamefont {Yan},\ and\ \citenamefont
  {Brady}}]{takatori2014prl}%
  \BibitemOpen
  \bibfield  {author} {\bibinfo {author} {\bibfnamefont {S.~C.}\ \bibnamefont
  {Takatori}}, \bibinfo {author} {\bibfnamefont {W.}~\bibnamefont {Yan}}, \
  and\ \bibinfo {author} {\bibfnamefont {J.~F.}\ \bibnamefont {Brady}},\ }\href
  {\doibase 10.1103/PhysRevLett.113.028103} {\bibfield  {journal} {\bibinfo
  {journal} {Phys.~Rev.~Lett.}\ }\textbf {\bibinfo {volume} {113}},\ \bibinfo
  {pages} {028103} (\bibinfo {year} {2014})}\BibitemShut {NoStop}%
\bibitem [{\citenamefont {Ginot}\ \emph {et~al.}(2015)\citenamefont {Ginot},
  \citenamefont {Theurkauff}, \citenamefont {Levis}, \citenamefont {Ybert},
  \citenamefont {Bocquet}, \citenamefont {Berthier},\ and\ \citenamefont
  {Cottin-Bizonne}}]{ginot2015prx}%
  \BibitemOpen
  \bibfield  {author} {\bibinfo {author} {\bibfnamefont {F.}~\bibnamefont
  {Ginot}}, \bibinfo {author} {\bibfnamefont {I.}~\bibnamefont {Theurkauff}},
  \bibinfo {author} {\bibfnamefont {D.}~\bibnamefont {Levis}}, \bibinfo
  {author} {\bibfnamefont {C.}~\bibnamefont {Ybert}}, \bibinfo {author}
  {\bibfnamefont {L.}~\bibnamefont {Bocquet}}, \bibinfo {author} {\bibfnamefont
  {L.}~\bibnamefont {Berthier}}, \ and\ \bibinfo {author} {\bibfnamefont
  {C.}~\bibnamefont {Cottin-Bizonne}},\ }\href {\doibase
  10.1103/PhysRevX.5.011004} {\bibfield  {journal} {\bibinfo  {journal} {Phys.
  Rev. X}\ }\textbf {\bibinfo {volume} {5}},\ \bibinfo {pages} {011004}
  (\bibinfo {year} {2015})}\BibitemShut {NoStop}%
\bibitem [{\citenamefont {Solon}\ \emph {et~al.}(2014)\citenamefont {Solon},
  \citenamefont {Fily}, \citenamefont {Baskaran}, \citenamefont {Cates},
  \citenamefont {Kafri}, \citenamefont {Kardar},\ and\ \citenamefont
  {Tailleur}}]{solon2014pressure}%
  \BibitemOpen
  \bibfield  {author} {\bibinfo {author} {\bibfnamefont {A.}~\bibnamefont
  {Solon}}, \bibinfo {author} {\bibfnamefont {Y.}~\bibnamefont {Fily}},
  \bibinfo {author} {\bibfnamefont {A.}~\bibnamefont {Baskaran}}, \bibinfo
  {author} {\bibfnamefont {M.}~\bibnamefont {Cates}}, \bibinfo {author}
  {\bibfnamefont {Y.}~\bibnamefont {Kafri}}, \bibinfo {author} {\bibfnamefont
  {M.}~\bibnamefont {Kardar}}, \ and\ \bibinfo {author} {\bibfnamefont
  {J.}~\bibnamefont {Tailleur}},\ }\href@noop {} {\bibfield  {journal}
  {\bibinfo  {journal} {arXiv:1412.3952}\ } (\bibinfo {year}
  {2014})}\BibitemShut {NoStop}%
\bibitem [{\citenamefont {Mallory}\ \emph {et~al.}(2014)\citenamefont
  {Mallory}, \citenamefont {\ifmmode \check{S}\else
  \v{S}\fi{}ari\ifmmode~\acute{c}\else \'{c}\fi{}}, \citenamefont {Valeriani},\
  and\ \citenamefont {Cacciuto}}]{mallory2014pre}%
  \BibitemOpen
  \bibfield  {author} {\bibinfo {author} {\bibfnamefont {S.~A.}\ \bibnamefont
  {Mallory}}, \bibinfo {author} {\bibfnamefont {A.}~\bibnamefont {\ifmmode
  \check{S}\else \v{S}\fi{}ari\ifmmode~\acute{c}\else \'{c}\fi{}}}, \bibinfo
  {author} {\bibfnamefont {C.}~\bibnamefont {Valeriani}}, \ and\ \bibinfo
  {author} {\bibfnamefont {A.}~\bibnamefont {Cacciuto}},\ }\href {\doibase
  10.1103/PhysRevE.89.052303} {\bibfield  {journal} {\bibinfo  {journal} {Phys.
  Rev. E}\ }\textbf {\bibinfo {volume} {89}},\ \bibinfo {pages} {052303}
  (\bibinfo {year} {2014})}\BibitemShut {NoStop}%
\bibitem [{\citenamefont {Risken}(1984)}]{risken1984fokker}%
  \BibitemOpen
  \bibfield  {author} {\bibinfo {author} {\bibfnamefont {H.}~\bibnamefont
  {Risken}},\ }\href@noop {} {\emph {\bibinfo {title} {Fokker-Planck
  Equation}}}\ (\bibinfo  {publisher} {Springer},\ \bibinfo {year}
  {1984})\BibitemShut {NoStop}%
\bibitem [{\citenamefont {Brown}\ \emph {et~al.}(2014)\citenamefont {Brown},
  \citenamefont {Vladescu}, \citenamefont {Dawson}, \citenamefont {Vissers},
  \citenamefont {Schwarz-Linek}, \citenamefont {Lintuvuori},\ and\
  \citenamefont {Poon}}]{brown2014swimming}%
  \BibitemOpen
  \bibfield  {author} {\bibinfo {author} {\bibfnamefont {A.~T.}\ \bibnamefont
  {Brown}}, \bibinfo {author} {\bibfnamefont {I.~D.}\ \bibnamefont {Vladescu}},
  \bibinfo {author} {\bibfnamefont {A.}~\bibnamefont {Dawson}}, \bibinfo
  {author} {\bibfnamefont {T.}~\bibnamefont {Vissers}}, \bibinfo {author}
  {\bibfnamefont {J.}~\bibnamefont {Schwarz-Linek}}, \bibinfo {author}
  {\bibfnamefont {J.~S.}\ \bibnamefont {Lintuvuori}}, \ and\ \bibinfo {author}
  {\bibfnamefont {W.~C.}\ \bibnamefont {Poon}},\ }\href@noop {} {\bibfield
  {journal} {\bibinfo  {journal} {arXiv:1411.6847}\ } (\bibinfo {year}
  {2014})}\BibitemShut {NoStop}%
\end{thebibliography}

%

\end{document}